# Electrical Spin Injection in a Ferromagnetic / Tunnel Barrier/ Semiconductor Heterostructure.

V.F. Motsnyi[1,2], V.I. Safarov[3], J. De Boeck[1], J. Das[1], W. Van Roy[1], E. Goovaerts[2] & G. Borghs[1]

[1] IMEC, Kapeldreef 75, B-3001 Leuven, Belgium

[2] University of Antwerp-UIA, Universiteitsplein 1, B-2610 Antwerpen, Belgium

[3] GPEC, case 901, Département de Physique, Faculté des Sciences de Luminy, 13288 Marseille, FRANCE

We demonstrate experimentally the electrical ballistic electron spin injection from a ferromagnetic metal / tunnel barrier contact into a semiconductor III-V heterostructure. We introduce the Oblique Hanle Effect technique for reliable optical measurement of the degree of injected spin polarization. In a CoFe / $Al_2O_3$ / GaAs / (Al,Ga)As heterostructure we observed injected spin polarization in excess of 8 % at 80K.

The functionality of spintronic devices [1] relies on the use of carrier spin rather than its charge. The development of spintronic devices is triggered by recent experiments showing that an electron spin polarized ensemble created inside a semiconductor, e.g. by circularly polarized light [2, 3], can have long drift lengths [4] and lifetimes [5] and can traverse semiconductor/semiconductor interfaces [6]. One of the main challenges for device implementation is the efficient creation of the spin-polarized ensemble within a semiconductor by electrical injection from a magnetic contact, spin injection. Spin injection from a diffusive ferromagnetic contact to a semiconductor has fundamental limitations [7, 8]. Magnetic semiconductors such as $Be_xMn_yZn_{1-x-y}Se$ [9] and GaMnAs [10] can inject spins into a semiconductor through diffusive ohmic contact only at low temperature and, in the former case, under large magnetic field. Tunnel injectors using ferromagnetic tips [11, 12] can work at room temperature but are not practical in devices. A Fe/GaAs Schottky barrier injector was recently reported to show spin-injection at RT [13]. Due to the interdiffusion and magnetic dead layer formation, a direct metal/semiconductor interface may lead to unreliable device operation. Our approach is to use the well-established $Al_2O_3$ insulator as a tunnel barrier between the ferromagnetic contact and the semiconductor heterostructure, a choice also supported by recent theoretical work [14]. We introduce the Oblique Hanle Effect (OHE) technique to assess spin polarized transport on a ferromagnet / semiconductor interface, a technique first used for optical detection of nuclear polarization [15].

The devices used in our experiments, shown in Fig.1, consist of a ferromagnetic contact, an $Al_2O_3$ tunnel barrier and an (Al)GaAs semiconductor heterostructure. Two different semiconductor heterostructures were grown by Molecular Beam Epitaxy on a (001) p-GaAs substrate. Sample 1 (Fig.1a): 2 μm p-GaAs layer ($p=2\times10^{18}$ cm$^{-3}$) / 200 nm p-$Al_{0.30}Ga_{0.70}As$ ($p-2\times10^{18}$ cm$^{-3}$) / 100 nm p-GaAs layer ($p=2\times10^{18}$ cm$^{-3}$), and Sample 2 (Fig.1b): 200 nm p-$Al_{0.30}Ga_{0.70}As$ ($p-2\times10^{18}$ cm$^{-3}$) / 100 nm GaAs (undoped) / 15 nm $Al_{0.20}Ga_{0.80}As$ (undoped). Both samples have an identical tunnel barrier contact: a 1.4 nm Al layer (oxidized in a controlled $O_2$ atmosphere forming a thin $Al_2O_3$ tunnel barrier) and the 2 nm $Co_{90}Fe_{10}$ / 8 nm $Ni_{80}Fe_{20}$ / 5 nm Cu magnetic stack. All metals are dc-magnetron sputtered. The ferromagnetic film magnetization is in-plane, showing a square hysteresis loop in the easy magnetization direction with a coercivity of about 0.4 kA/m. Surface emitting LED structures were fabricated with magnetic contact sizes of 20×20 μm$^2$ using optical lithography, dry and wet processing steps. The devices were contacted using Au contacts to the backside of the substrate and to the ferromagnetic contact leaving an optical window on the perimeter. Under forward bias conditions, light emission corresponding to GaAs band gap transitions occurs. At ~80 K the light emission threshold is about 1.4 V and 1.6 V for Sample 1 and Sample 2, respectively. In order to get sufficient signal to noise ratio, the measurements were carried out at about 2 V and 3 V biasing for Sample 1 and Sample 2, respectively, with typical current around 90 mA.

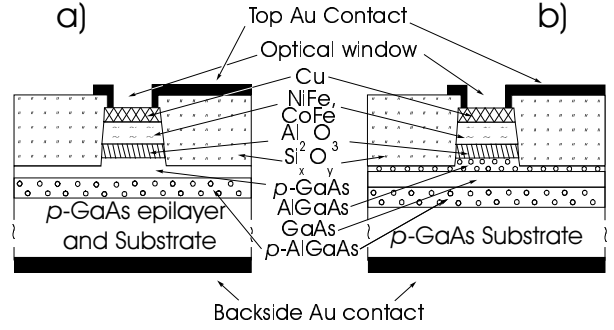

*Fig.1. Cross sectional view of the samples for electrical spin injection from a magnetic tunnel barrier. (See text for further details).*

Circular polarization of the electroluminescence is the natural way to measure spin orientation of injected electrons in III-V semiconductors [2]. Due to the high refractive index of GaAs, the degree of circular polarization of the emitted light characterizes only the component of the electron spin normal to the surface. With spin injection from the in-plane magnetized ferromagnet, no circular polarization of a surface emitting LED structure is expected, unless the spins are manipulated to obtain a non-zero perpendicular spin component. We propose to use the Oblique Hanle Effect (OHE) technique, described below, as a powerful tool to assess the spin injection.

The electron spin population is characterized by the average electron spin $\vec{S} = \left(\sum_{i=1}^{n}\vec{s}_i\right)/n$, where $\vec{s}_i$ is the spin of the individual electrons, n is the number of electrons. The degree of spin polarization Π along the direction of average electron spin, is $\Pi = (n^\uparrow - n^\downarrow)/(n^\uparrow + n^\downarrow) = 2\cdot|\vec{S}|$. In the presence of an external magnetic field $\vec{B}$ ($|\vec{B}\times\vec{S}| \neq 0$) the average spin undergoes precession with the Larmor frequency $\vec{\Omega} = \frac{g^* \cdot \mu_b}{\hbar}\cdot\vec{B}$, were $g^*$- is the effective g-factor and $\mu_B$ the Bohr magneton. The evolution of the average spin momentum $\vec{S}$ in the semiconductor in an external magnetic field, taking into account the spin relaxation and injection/recombination processes is described by [2]:

$$\frac{d\vec{S}}{dt} = \frac{\vec{S}_O}{\tau} - \frac{\vec{S}}{T_S} + \vec{\Omega}\times\vec{S} \qquad (1)$$

where $\vec{S}_O$ is the average injected spin momentum, $\tau$ the lifetime of electrons and $T_S$ is the spin lifetime ($T_S^{-1} = \tau^{-1} + \tau_S^{-1}$), $\tau_S$ is the spin relaxation time. Let us assume the following geometry: the xy-plane describes the sample surface, the easy axis of magnetization points along the x-axis and the magnetic field lies in the zx-plane with an angle $\varphi$ to the surface normal. Under steady state conditions ($d\vec{S}/dt = 0$) the solution of Eq.(1) can be easily found. By applying an external magnetic field, starting from an injected average spin $\vec{S}_O(S_{Ox},0,0)$, spin precession leads to a non-zero $S_Z$-component giving rise to circular polarization of the light propagating along the z-direction. From Eq.(1), $S_Z$ has the following dependency on magnetic field:

$$S_z(B) = S_{Ox} \cdot \frac{T_s}{\tau} \cdot \frac{(\frac{g^*\mu_b}{\hbar}BT_s)^2 \cdot \cos\varphi \cdot \sin\varphi}{1+(\frac{g^*\mu_b}{\hbar}BT_s)^2} \quad (2)$$

It grows from zero and saturates at higher values of external magnetic field. The saturation value is dependent on the magnetic field orientation ($\varphi$), reaching a maximum of $S_{Z\max} = \frac{1}{2} \cdot S_{Ox} \cdot \frac{T_S}{\tau}$ for $\varphi = \pi/4$. The selection rules for interband optical transitions give the simple relation between the degree of circular polarization $P = (I^{\sigma^+} - I^{\sigma^-})/(I^{\sigma^+} + I^{\sigma^-})$, where $I^{\sigma^+}$ and $I^{\sigma^-}$ are intensities of right and left circular polarized radiation, and the average electron spin component $P(B)=S_Z(B)$. Hence by measuring $P(B)$, the average injected electron spin $S_{Ox}$ and the degree of spin polarization $\Pi = 2\cdot S_{Ox} = 4\cdot S_{Z\max}\frac{\tau}{T_s}$ can be determined if $T_S$ and $\tau$ are known. Experimental measurements of $T_S$ and $\tau$ can be performed by OHE using *optical spin injection* ($\vec{S}_O(0,0,S_{Oz}=1/4)$) [2], and detection [16]. In the same geometry, due to the precession around the increasing oblique magnetic field, $S_Z$ is maximal at $B=0$ and decreases at higher values of magnetic field. From Eq.(1), $S_Z$ in the case of *optical spin injection* has the following dependency on magnetic field:

$$S_z(B) = S_{Oz} \cdot \frac{T_s}{\tau} \cdot \frac{1+(\frac{g^*\mu_b}{\hbar}BT_s)^2 \cdot \cos^2\varphi}{1+(\frac{g^*\mu_b}{\hbar}BT_s)^2} \quad (3)$$

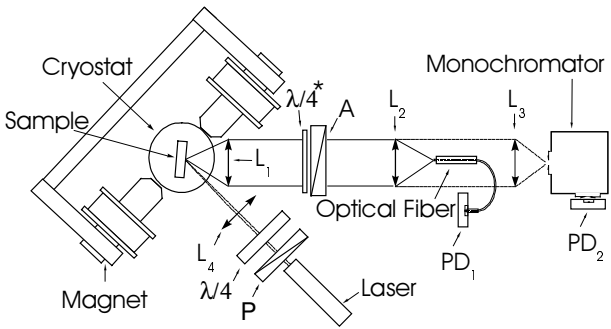

*Fig.2. Schematic view on the experimental setup for the investigation of electrical and optical spin injection into a semiconductor using the Oblique Hanle Effect (OHE) technique.*

The experimental set-up, shown in Fig.2, allows to measure the degree of circular polarization of the luminescence using electrical and optical excitation under oblique magnetic field. An optical nitrogen cryostat allows the cooling of the sample to about 80 K. The magnet provides an external magnetic field up to 0.6 T. The emitted light is coupled into an optical fiber by lens $L_2$ and is detected by a photodetector ($PD_1$). A rotating quarter-wave plate ($\lambda/4^*$) and analyzer (A) together with lock-in detection allow precise measurements of the degree of circular polarization of the emitted light. For optical spin injection and detection, semiconductor ($h\nu$=1.58 eV) and He-Ne ($h\nu$=1.96 eV) lasers (L) together with an optical monochromator and photodetector $PD_2$ are used.

In Fig.3.a we first present the OHE-measurements using optical spin injection and detection to determine $T_S$ and $\tau$ for sample 1. The open circles correspond to the measurements of circular polarization of the light with optical excitation $h\nu$=1.58 eV, i.e., near the GaAs band gap. The solid line represents the fit using Eq.(3) with the following parameters: $T_s/\tau$ =0.48, $T_s$ =0.12 ns. An identical experiment with photons of higher energy $h\nu$=1.96 eV (Fig.3a, diamonds) shows the same sign of circular polarization, indicating [2, 17] that the electrons keep their spin orientation during the thermalization to the bottom of the conduction band. The reduction of the degree of polarization is due to the fact that "cold" electrons excited from the split-off valence band have the opposite spin orientation.

This result is now used to interpret the electrical OHE spin injection measurements using the magnetic tunnel contact shown in Fig.3b (curve (i)). The measured curve does not fit the Eq.(2) predicted for the OHE, due to the superposition of the spin injection signal and a magnetooptical effect in the ferromagnetic film. In an oblique magnetic field the Magnetic Circular Dichroism (MCD, difference in transmission of right hand and left hand circularly polarized light) occurs, which can be measured in a photoluminescence experiment on the same set up with *linearly polarized* excitation, $\Pi_{inj}$=2·$S_Z$=0 [2]. Curve (ii) on Fig.3b shows the resulting circular polarization in the luminescence with a linear dependence on the oblique field, as expected for the MCD and the experimental configuration. (Here we neglect the contribution of MCD on the polarization of optically excited electrons, for the following reason. If one would imagine that curve (i) on Fig.3b is caused by MCD then linearly polarized light will obtain 0.33% circular polarization after propagation through the FM film. Selection rules under absorption and emission, as well as measured $T_S/\tau$ ratio (Fig.3a) will result in the emission of light with only 0.04% circular polarization.)

The difference between the measured degree of circular polarization by electrical injection and that caused by MCD (curves (i) and (ii) on Fig.3b) gives the net effect of the injected spin polarization (Fig.3c), which now can be fit (solid line) using Hanle curve, Eq.(2), with the following parameters: $T_s/\tau$ =0.48, $T_s$ =0.12 ns and $S_{Ox}$=0.0062. The *only* fit parameter is $S_{Ox}$, the average spin momentum of electronically injected electrons. All other parameters were taken from the measurements of OHE with optical spin injection. The degree of spin polarization $\Pi$ of electrically injected electrons was found to be 1,2±0.1%.

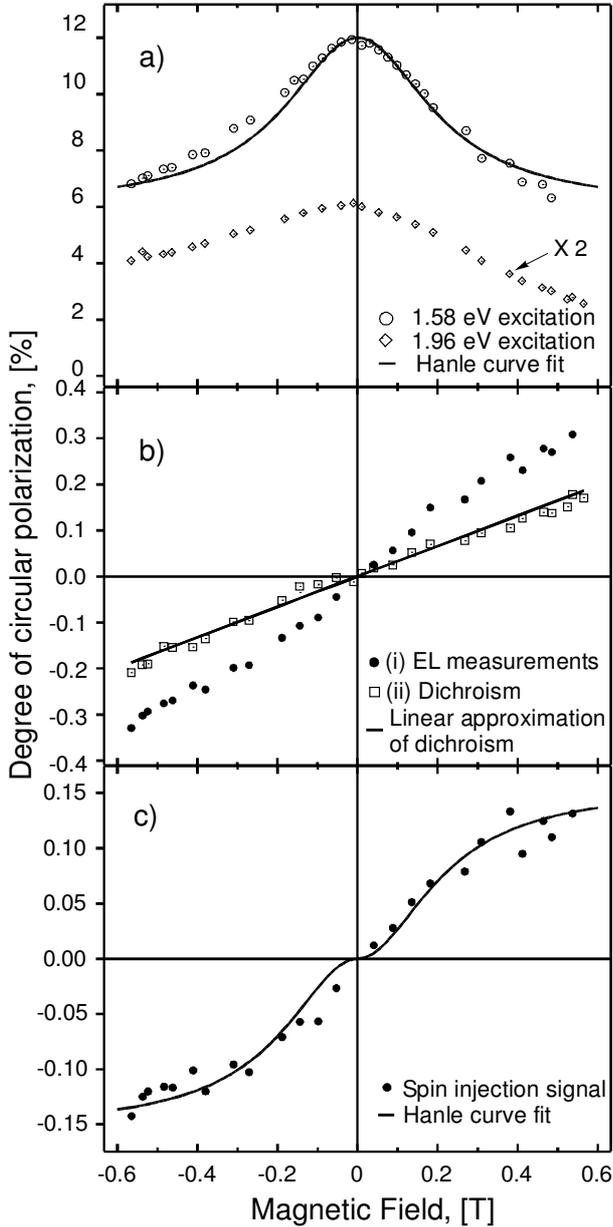

*Fig.3. Set of OHE measurements with optical and electrical spin injection on sample 1 (Fig.1a) at T~80 K. a) Damping of the degree of circular polarization in the oblique magnetic field ($\varphi=\pi/4$) under optical spin injection. b) Measured degree of circular polarization of the electroluminescence (EL) under (i) electrical spin injection using OHE technique, and (ii)- optical linearly polarized laser excitation with $h\nu=1.58\,eV$ ($\varphi=\pi/4$). c) Electrical spin injection signal (difference between curves (i) and (ii) of Fig.3b), and the fit to the Hanle curve, Eq.2. The change of the sign of the circular polarization is caused by the switching of the magnetic contact by the external magnetic field.*

In an effort to study the effect of a different $Al_2O_3$ / semiconductor interface, we fabricated Sample 2 (Fig.1b), comprising an AlGaAs (15nm) undoped interface layer. The OHE measurement with electrical injection is shown in Fig 4. In this case the component of circular polarization caused by MCD, as measured in Sample 1, is small in comparison with the measured circular polarization and fits nicely to the Hanle curve

Eq.(2) (solid line) with following parameters: $S_{Ox} \cdot \frac{T_s}{\tau} = 0.041$ and $T_s = 0.2$ ns. In this case the value $\left(\Pi \cdot \frac{T_s}{\tau}\right)$ is found to be 8.2±0.2%. The values for $T_S$ and $\tau$ are not known for sample 2, since in the supporting experiment, OHE with optical spin injection, the photoluminescence is dominated by emission from the highly doped GaAs substrate. The 8.2% is a lower limit for electron spin polarization $\Pi$, since $T_S/\tau = \tau_S/(\tau_S + \tau) \leq 1$.

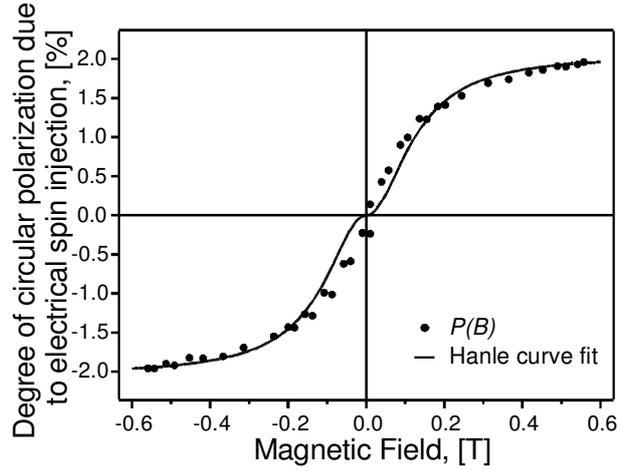

*Fig.4. Measured degree of circular polarization of the light on sample 2 (Fig.1b) at T~80 K under electrical spin injection using the OHE technique ($\varphi=\pi/4$) and the fit to the Hanle curve, Eq.(2). The change of the sign of the circular polarization is caused by the switching of the magnetic contact by the external magnetic field.*

The strong decrease of the quantum efficiency of the LED and the $\frac{T_s}{\tau}$ ratio with increasing temperature does not allow reliable electrical OHE measurements at room temperature at present. The spin injection in the present devices can be increased the same way as in the Tunnel Magnetoresistance Junctions (TMR), by improving the quality of the oxide barrier, its interfaces and by reducing the bias across the tunnel barrier [18,19,20]. Our results show that the Oblique Hanle Effect is a powerful tool to assess spin injection. The tunneling process is temperature independent, so that at room temperature the same value of spin injection should occur. These are promising results for future room temperature spintronic devices using stable tunnel barrier injectors in III-V (e.g.GaAs, GaN) or state-of-the-art $SiO_2$/silicon technology.

We thank Hans Boeve for help in tunnel barrier fabrication and ferromagnetic metal deposition, Willem van de Graaf for MBE sample growth, Liesbet Lagae for experimental assistance, Reiner Windisch and Barun Dutta for experimental assistance and discussions. JD acknowledges financial support from the I.W.T. (Belgium). WVR acknowledges financial support as a Postdoctoral Fellow of the Fund for Scientific Research Flanders-Belgium (F.W.O.).